\documentclass{PoS}

\usepackage{amsmath}

\newcommand{\lp}{\left(}
\newcommand{\rp}{\right)}

\title{Gauge coupling beta functions in the Standard Model}

\ShortTitle{Gauge coupling beta functions in the Standard Model}

\author{Luminita Mihaila, 
  Jens Salomon, 
  \speaker{Matthias Steinhauser}%
  \\
  KIT\\
  E-mail: \email{matthias.steinhauser@kit.edu}}

\abstract{We report about the computation of three-loop
  corrections to the gauge coupling beta functions in the Standard Model.}

\FullConference{Loops and Legs in Quantum Field Theory\\
  11th DESY Workshop on Elementary Particle Physics \\
  April 15-20, 2012\\
  Wernigerode, Germany}

\begin{document}

\section{Introduction}

Renormalization group functions play an important role in quantum field
theory. They determine the energy scale dependence of the parameters of the
Lagrange density and thus are important tools to combine predictions of the
theory from different energy regions. An important example in this respect is
the inspection of the gauge coupling unification at high energies where
precise experimental data at the electroweak scale combined
with accurate calculations of the renormalization constants yields
precise predictions.

In this contribution we consider the renormalization group functions of the
three gauge coup\-lings in the Standard Model (SM) within the Modified Minimal
Subtraction ($\overline{\rm MS}$) scheme. In this renormalization scheme the
beta functions are independent of all
mass scales present in the theory and it is thus relatively simple to solve
the loop integrals. In fact, in the calculation presented here at most 
massless three-loop two-point functions have to be evaluated which are known
for more than 30 years~\cite{Chetyrkin:1981qh}.
Actually the main difficulty of the calculation is the huge amount of
contributing Feynman diagrams (about $10^6$)
which result from the large number of vertices and
propagators. This requires an automated setup which not only generates and
computes all Feynman diagrams but also provides the Feynman rules in an
automated way.

The results presented in these proceedings have been obtained in
Refs.~\cite{Mihaila:2012fm,MSS_2}. There have been a number of publications
where the one- and two-loop expressions have been
computed~\cite{Gross:1973id,Politzer:1973fx,Jones:1974mm,Tarasov:1976ef,Caswell:1974gg,Egorian:1978zx,Jones:1981we,Fischler:1981is,Machacek:1983tz,Jack:1984vj}.
Also several three-loop results have been computed since the end of the
seventies~\cite{Curtright:1979mg,Jones:1980fx,Tarasov:1980au,Larin:1993tp,Steinhauser:1998cm,Pickering:2001aq}. Four-loop
corrections to beta functions are only known for
QCD~\cite{vanRitbergen:1997va,Czakon:2004bu}.

Let us in a first step define the couplings we want to consider. It is
convenient to use instead of the fine structure constant $\alpha_{\rm QED}$
and the weak mixing angle $\theta_W$ the gauge couplings in a SU(5)-like
normalization given by
\begin{align}
  \alpha_1 = \frac{5}{3}\frac{\alpha_{\rm QED}}{\cos^2\theta_W}\,,\qquad
  \alpha_2 = \frac{\alpha_{\rm QED}}{\sin^2\theta_W}\,,\qquad
  \alpha_3 = \alpha_s\,.
  \label{eq::alpha_123}
\end{align}
Note that it is straightforward to obtain the beta functions for $\alpha_{\rm
  QED}$ and $\sin^2\theta_W$ once the $\beta_{\alpha_i}$ are know (see, e.g.,
Ref.~\cite{MSS_2}).

\section{Calculation}

We define the beta functions via
\begin{equation}
  \mu^2\frac{d}{d\mu^2}\frac{\alpha_i(\mu)}{\pi}
  =\beta_i(\{\alpha_j(\mu)\},\epsilon)\,,
\label{eq::beta_fc}
\end{equation}
where $i=1,2,3$ labels the three gauge couplings.  The index $j$ runs over all
couplings in the SM, i.e., the gauge, Yukawa and Higgs boson self couplings. We
furthermore have $\epsilon=(4-d)/2$ where $d$ is the space-time dimension used
for the evaluation of the momentum integrals.

The functions $\beta_i(\{\alpha_j(\mu)\},\epsilon)$ are conveniently computed
from the renormalization constants relating the bare and renormalized gauge
couplings via 
\begin{eqnarray}
  \alpha_i^{\rm bare} &=&
  \mu^{2\epsilon}Z_{\alpha_i}(\{\alpha_j\},\epsilon)\alpha_i
  \,. 
\end{eqnarray}
Inserting this
equation into~(\ref{eq::beta_fc}) and exploiting the fact that
$\alpha_i^{\rm bare}$ does not depend on $\mu$ leads to
\begin{eqnarray}
  \label{eq::renconst_beta}
  \beta_i &=& 
  -\left[\epsilon\frac{\alpha_i}{\pi}
    +\frac{\alpha_i}{Z_{\alpha_i}}
    \sum_{{j=1},{j \neq i}}
    \frac{\partial Z_{\alpha_i}}{\partial \alpha_j}\beta_j\right]
  \left(1+\frac{\alpha_i}{Z_{\alpha_i}}
    \frac{\partial Z_{\alpha_i}}{\partial \alpha_i}\right)^{-1}
  \,,
\end{eqnarray}
which constitutes the master formula for our enterprise.  From this equation
it is clear that it is sufficient to compute the renormalization constants
$Z_{\alpha_j}$ in the $\overline{\rm MS}$ scheme in order to obtain
$\beta_i$. In fact, we have to compute $Z_{\alpha_1}$, $Z_{\alpha_2}$ and
$Z_{\alpha_3}$ to three-loop order and the renormalization constants for the Yukawa
coupling to one-loop order.  As far as the Higgs boson self coupling is
concerned it is sufficient to have the leading term proportional to $\epsilon$
of the corresponding beta function.  The discussion in the following is
centered around the three-loop calculation of the gauge coupling
renormalization constants.

The procedure for the calculation of 
$Z_{\alpha_i}$ ($i=1,2,3$) is as follows: (i) choose a vertex which contains 
the coupling $\alpha_i$;  (ii) compute the renormalization constant of that
vertex, $Z_{\rm vrtx}$; (iii) compute the wave function renormalization
constant of the external particles, $Z_{k,{\rm wf}}$ ; and (iv) combine them
according to $Z_{\alpha_i} = (Z_{\rm{vrtx}})^2/(\prod_k Z_{k,{\rm{wf}}})$. 

We have used two independent approaches to compute the beta functions. The
first one is based on the formulation of the SM using Lorenz gauge in the
unbroken phase, i.e., all particles are still massless. Since the beta
functions are mass independent this setup is quite convenient as the structure
is simpler than after spontaneous symmetry breaking.  The gauge bosons
in this approach are the $B$ and $W$ bosons and the gluons.

\begin{figure}
  \begin{center}
    \includegraphics[width=.95\textwidth]{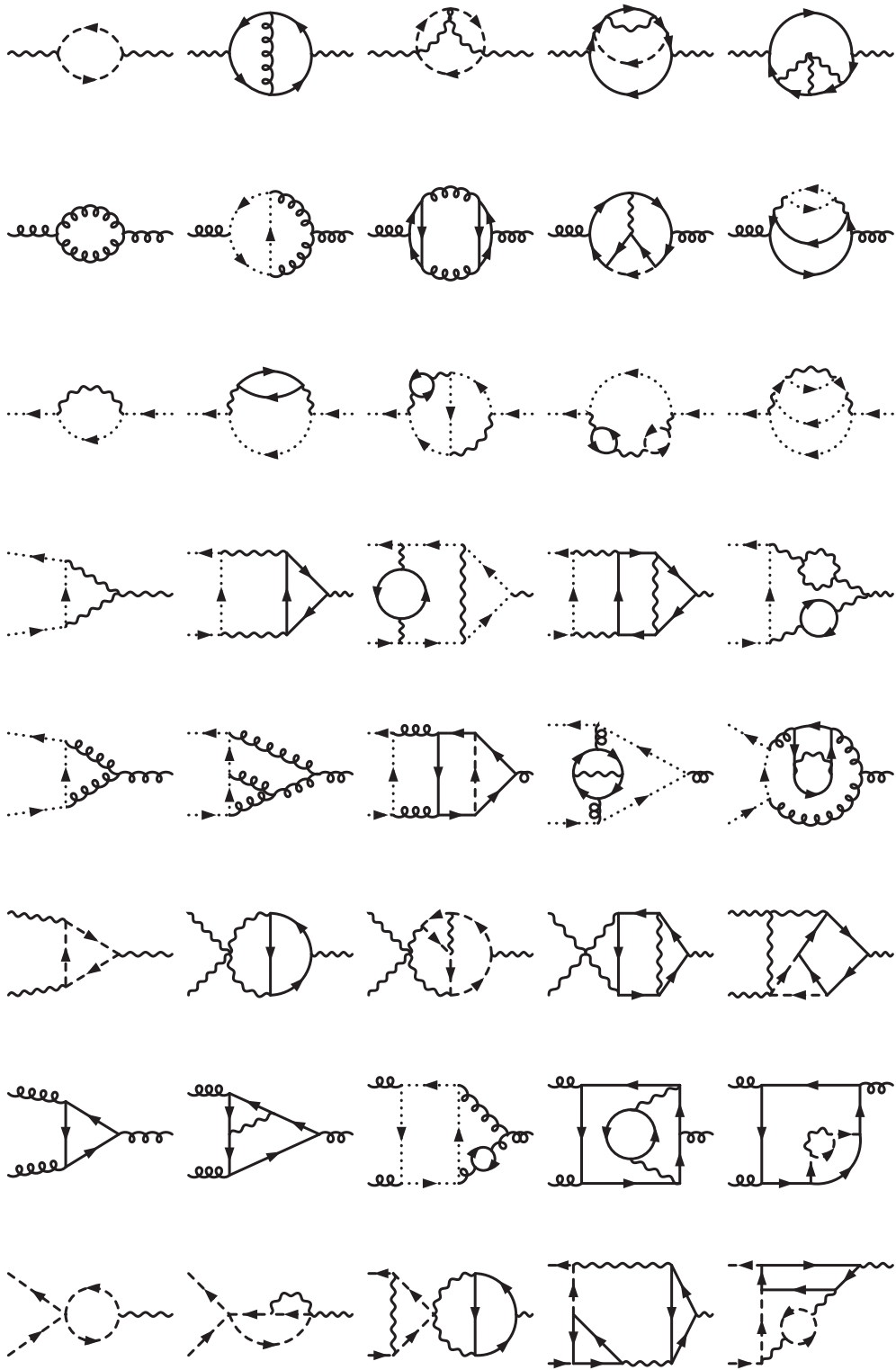}
    \caption{\label{fig::diags}Sample Feynman diagrams contributing to 
      the Green's functions which have been used for the calculation of the 
      renormalization constants of the gauge couplings.
      Solid, dashed, dotted, curly and wavy lines correspond to fermions,
      Higgs bosons, ghosts, gluons and electroweak gauge bosons,
      respectively.} 
  \end{center}
\end{figure}

For the computation of $Z_{\alpha_2}$ and $Z_{\alpha_3}$ it is convenient to
choose the gauge boson ghost vertex since in that case the number of
contributing diagrams is smallest. In order to have a cross check we have
chosen the triple gauge boson vertex as a second option.  For $Z_{\alpha_2}$
also the $\phi^+ \phi^- W_3$ vertex as a third alternative has been used.
Note that due to the Ward identity $Z_{\alpha_1}$ is solely obtained from the
$B$ boson two-point function.  Sample Feynman diagrams for all two- and
three-point Green's functions can be found in Fig.~\ref{fig::diags}.

In the second approach we have used the SM Lagrangian in the background field
gauge (BFG) as a starting point (see, e.g., Ref.~\cite{Denner:1994xt}). The
advantage of this method is that only the gauge boson two-point functions and
thus less different Green's functions have to be considered. On the other hand
more vertices are present and thus more Feynman diagrams contribute to the
background gauge boson propagator as compared to the corresponding quantum
version. We use the BFG in the broken phase of the SM. However, here
we can also set the masses of all particles to zero.  Sample Feynman diagrams for
the gauge boson two-point functions are shown in the first and second row of
Fig.~\ref{fig::diags} where the external lines correspond to background gluon,
photon and $W$ or $Z$ bosons.

An important issue for all loop calculations involving electroweak gauge
bosons is the treatment of $\gamma^5$ in $d\not=4$ dimensions.  In our
calculation we on purpose do not use Green's functions involving external
fermions. For all other two- and three-point functions we could show that a
naive treatment of $\gamma^5$ leads to the correct result~\cite{MSS_2}.  Note
that Green's functions with external fermions are unavoidable for the
calculation of the beta functions for the Yukawa couplings. At two-loop order
it is again possible to use a naive version of $\gamma_5$. Beyond two loops,
however, a more careful treatment is required (see, e.g.,
Ref.~\cite{Chetyrkin:2012rz}).

We have performed several checks which convinced us from the correctness of our
results. In brief they are given by:
\begin{itemize}
\item comparison of the one- and two-loop results with the literature
\item comparison of partial three-loop results with the literature
\item computation of the beta function for the Higgs boson self-coupling to
  one-loop order and comparison with the literature
\item computation of the Yukawa beta functions for the top and
  bottom quark, and the tau lepton to two-loop order and comparison with
  the literature
\item computation of the $BBB$ vertex to three-loop order; we checked that 
  the sum of all 358\,716 diagrams gives zero
\item computation for general gauge parameters; we checked that they drop out in
  the final result for $\beta$ functions
\item check that no infra-red divergences are present in the loop integrals
\end{itemize}

As a last comment on the technical details of our calculation let us mention
that we were able to obtain the final result for the gauge coupling beta
functions for a general Yukawa structure involving all nine Yukawa couplings of
the SM and the CKM matrix in the quark sector. Furthermore, it is
straightforward to extend our final result for a SM with fourth family as will
be explained in the next section.

\section{Results}

In this section we briefly discuss the final results for the gauge coupling beta
functions.  In order to show the structure of the analytical expressions we
present the result for $\beta_1$ which is given by
\begin{align}
  \beta_1 &=
  \frac{\alpha_1^2}{\lp4\pi\rp^2} \bigg\{ \frac{2}{5} + \frac{16 n_G}{3} \bigg\} \notag \\
  &  + \frac{\alpha_1^2}{\lp4\pi\rp^3} \bigg\{ \frac{18 \alpha_1}{25} + \frac{18 \alpha_2}{5} - \frac{34 \text{tr}\hat{T}}{5} - 2 \text{tr}\hat{B} - 6 \text{tr}\hat{L} + n_G \bigg[ \frac{76 \alpha_1}{15} + \frac{12 \alpha_2}{5} + \frac{176 \alpha_3}{15} \bigg] \bigg\} \notag \\
  &  + \frac{\alpha_1^2}{\lp4\pi\rp^4} \bigg\{ \frac{489 \alpha_1^2}{2000} + \frac{783 \alpha_1 \alpha_2}{200} + \frac{3401 \alpha_2^2}{80} + \frac{54 \alpha_1 \hat{\lambda}}{25} + \frac{18 \alpha_2 \hat{\lambda}}{5} - \frac{36 \hat{\lambda}^2}{5} - \frac{2827 \alpha_1 \text{tr}\hat{T}}{200} \notag \\
  &  - \frac{471 \alpha_2 \text{tr}\hat{T}}{8} - \frac{116 \alpha_3 \text{tr}\hat{T}}{5} - \frac{1267 \alpha_1 \text{tr}\hat{B}}{200} - \frac{1311 \alpha_2 \text{tr}\hat{B}}{40} - \frac{68 \alpha_3 \text{tr}\hat{B}}{5} - \frac{2529 \alpha_1 \text{tr}\hat{L}}{200} \notag \\
  &  - \frac{1629 \alpha_2 \text{tr}\hat{L}}{40} + \frac{183 \text{tr}\hat{B}^2}{20} + \frac{51 (\text{tr}\hat{B})^2}{10} + \frac{157 \text{tr}\hat{B}\text{tr}\hat{L}}{5} + \frac{261 \text{tr}\hat{L}^2}{20} + \frac{99 (\text{tr}\hat{L})^2}{10} \notag \\
  &  + \frac{3 \text{tr}\hat{T}\hat{B}}{2} + \frac{339 \text{tr}\hat{T}^2}{20} + \frac{177 \text{tr}\hat{T}\text{tr}\hat{B}}{5} + \frac{199 \text{tr}\hat{T}\text{tr}\hat{L}}{5} + \frac{303 (\text{tr}\hat{T})^2}{10} \notag \\
  &  + n_G \bigg[ - \frac{232 \alpha_1^2}{75} - \frac{7 \alpha_1 \alpha_2}{25} + \frac{166 \alpha_2^2}{15} - \frac{548 \alpha_1 \alpha_3}{225} - \frac{4 \alpha_2 \alpha_3}{5} + \frac{1100 \alpha_3^2}{9} \bigg] \notag \\
  & + n_G^2 \bigg[ - \frac{836 \alpha_1^2}{135} - \frac{44 \alpha_2^2}{15} -
  \frac{1936 \alpha_3^2}{135} \bigg] \bigg\} \,,
\label{eq::beta1}
\end{align}
where the limit $\epsilon\to0$ has been taken.
For each gauge coupling beta function we have that the one-loop contribution of
$\beta_i$ is proportional to $\alpha_i^2$. Mixed contributions of order
$\alpha_i^2\alpha_j$ and $\alpha_i^2\alpha_k\alpha_l$ only appear
at two and three loops, respectively, where $\alpha_j$ 
are gauge or Yukawa and $\alpha_k$ and $\alpha_l$
are gauge, Yukawa or Higgs boson self couplings.
Note that the latter appears for the first time at three-loop order.

\begin{figure}
  \begin{center}
  \includegraphics[width=0.7\linewidth]{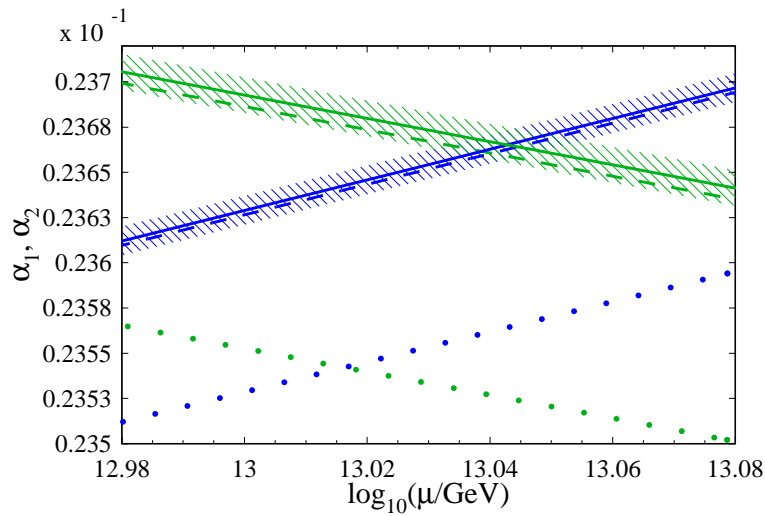}
  \caption{The running of the electroweak gauge couplings in the SM. The lines
    with positive slope correspond to $\alpha_1$, the lines with negative
    slope to $\alpha_2$.  The dotted, dashed and solid lines correspond to
    one-, two- and three-loop precision, respectively.  The bands around the
    three-loop curves visualize the experimental
    uncertainty.}\label{fig::run2}
  \end{center}
\end{figure}

The symbol $n_G$ in Eq.~(\ref{eq::beta1}) stands for the number of
generations; in the SM we have $n_G=3$.
The quantities $\text{tr}\hat{B}$, $\text{tr}\hat{T}$ and
$\text{tr}\hat{L}$ incorporate the Yukawa couplings where we refer to
Ref.~\cite{MSS_2} for details. In this contribution we only want to mention 
that the replacements
\begin{eqnarray}
  \text{tr}\hat{L}^n \to \alpha_{\tau}^n\,,\quad
  \text{tr}\hat{T}^n \to \alpha_{t}^n\,,\quad
  \text{tr}\hat{B}^n \to \alpha_{b}^n\,,\quad
  \text{tr}\hat{T}\hat{B} \to \alpha_{t}\alpha_{b}\,,
\end{eqnarray}
leads to the result where only the Yukawa coupling of the third generation is
kept non-zero.

In a similar way we can accommodate a fourth generation of fermions.  In this
case, the Yukawa matrices become $4\times 4$ dimensional.  If we assume that
the fourth is much heavier and if we neglect all SM Yukawa interactions it
contains a $3\times 3$ zero matrix and we have
\begin{equation}
  \hat{F_4} = \left(\begin{matrix}
      0_{3\times 3} & 0  \\
      0        & \alpha_F 
    \end{matrix}
  \right)\,,\quad \mbox{with} \quad F=T, B, L\,,
\end{equation}
where $T$ and $B$ stand for the up- and down-type heavy quarks, and $L$ for
the heavy charged and neutral leptons, while 
$\alpha_F = \frac{\alpha m_F^2}{2 \sin^2\theta_W M_W^2}$ 
denotes the corresponding Yukawa couplings.
Note that the contribution of a heavy neutrino is not contained in our
formulae.

Let us finally briefly discuss the numerical impact of the new three-loop 
corrections. In Fig.~\ref{fig::run2} we show the 
running of $\alpha_1$ and $\alpha_2$ from $\mu=M_Z$ to the energy scales 
where these two coupling become equal. The dotted and dashed lines correspond
to one- and two-loop running, respectively. One observes a significant change
of the curves, which is in particular much bigger than the experimental
uncertainty indicated by the dashed band. Thus in case only one- and two-loop
perturbative corrections are included the theory uncertainty is much bigger
than the experimental one. This changes with the inclusion of the three-loop
terms. The  results are shown as solid lines which are close to the
corresponding dashed curves. The effect is small, however, still of the order
of the experimental uncertainty, in particular for $\alpha_2$.

To conclude, the complete three-loop corrections to the gauge coupling beta
functions have been computed~\cite{Mihaila:2012fm,MSS_2} using 
different methods and applying several cross checks. 
They constitute fundamental quantities of the
SM and are important for high-precision experimental checks.

\end{document}